\documentclass[twocolumn,showpacs,preprintnumbers,amsmath,amssymb]{revtex4}

\usepackage{graphicx}
\usepackage{dcolumn}
\usepackage{bm}
\usepackage{ulem} 

\def\figpdf  {\includegraphics[width=8cm]{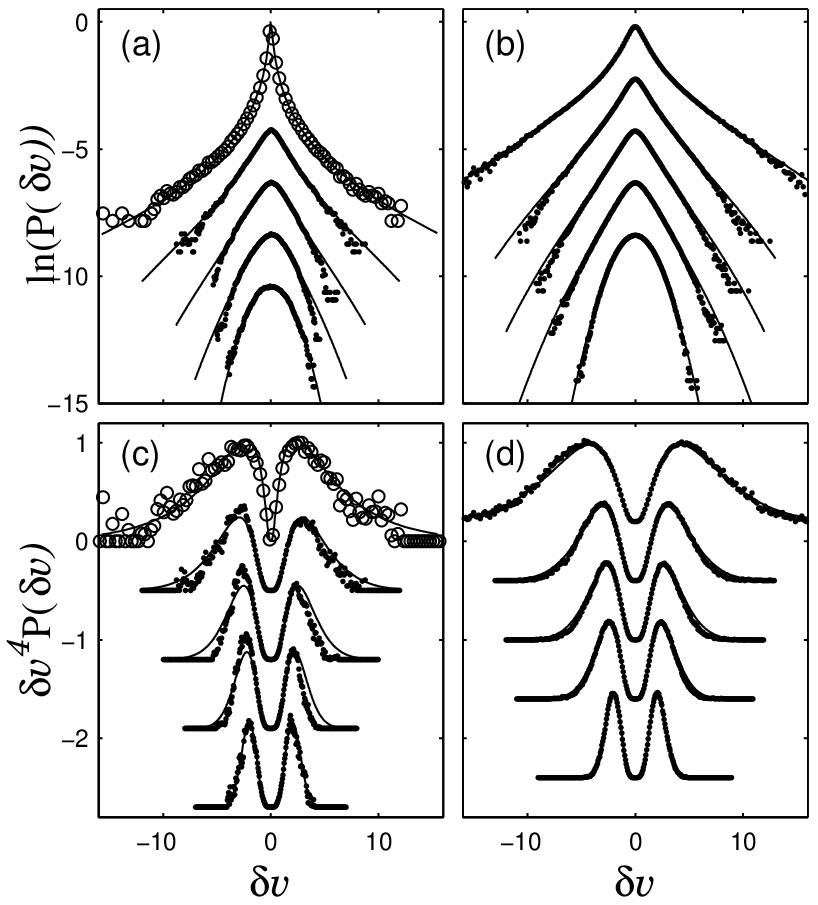}}
\def\figcum  {\includegraphics[width=8cm]{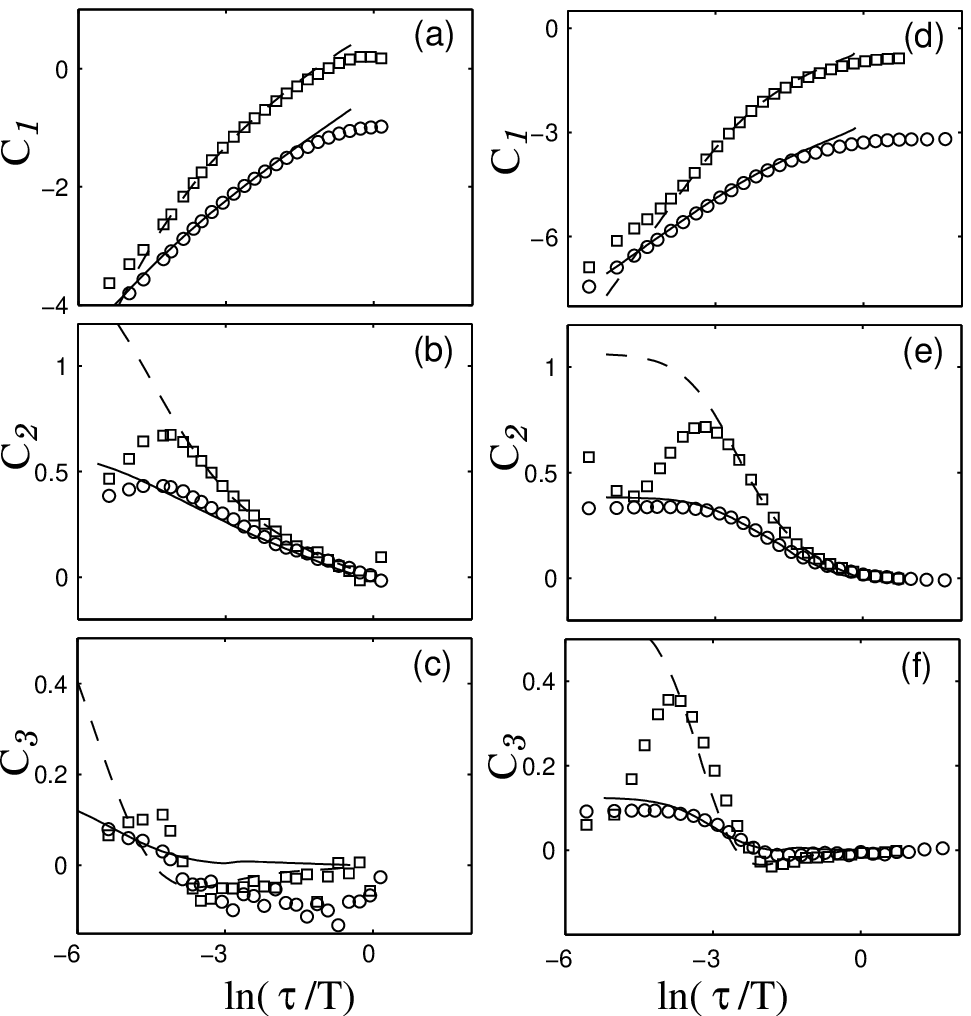}}
\def\figdh  {\includegraphics[width=6cm]{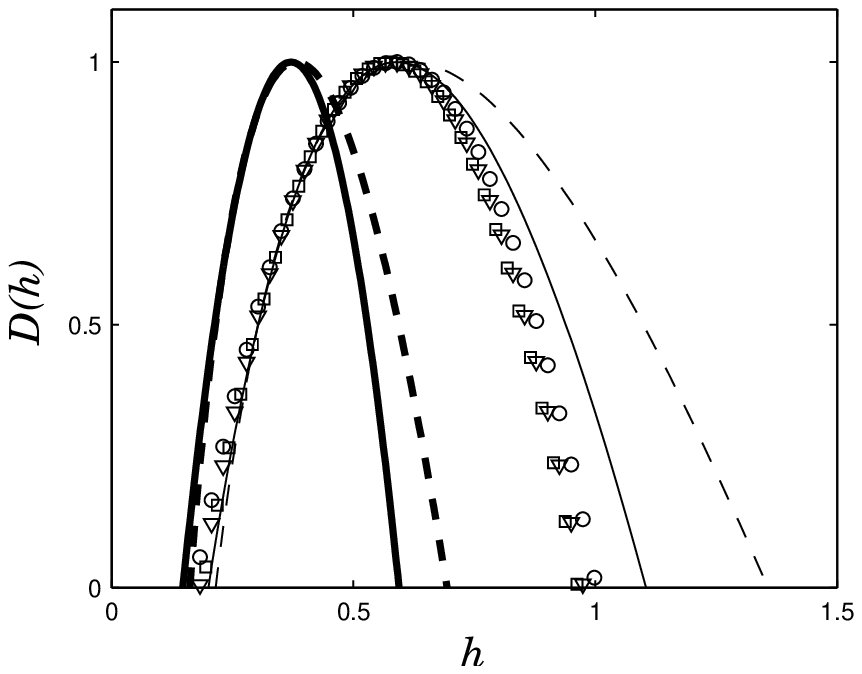}}

\begin{document}


\title{ Lagrangian Velocity Statistics in Turbulent Flows: Effects of Dissipation}
\author{L. Chevillard}
\author{S.G. Roux}
\author{E. Lev\^eque}
\author{N. Mordant}
\altaffiliation{Present address: Laboratory of Atomic State and Solid State Physics, Cornell University, Ithaca, New-York 14853-2501, USA}
\author{J.-F. Pinton}
\author{A. Arneodo}
\affiliation{
Laboratoire de Physique, \'Ecole Normale Sup\'erieure de Lyon\\
46 all\'ee d'Italie, F-69007 Lyon, France}

\date{\today}

\begin{abstract}
We use the multifractal formalism to describe the effects of dissipation on Lagrangian velocity statistics in turbulent flows. We analyze high Reynolds number experiments and direct numerical simulation (DNS) data. We show that this approach reproduces the shape evolution of velocity increment probability density functions (PDF) from Gaussian to stretched exponentials as the time lag decreases from integral to dissipative time scales. A quantitative understanding of the departure from scaling exhibited by the magnitude cumulants, early in the inertial range, is obtained with a free parameter function $D(h)$ which plays the role of the singularity spectrum in the asymptotic limit of infinite Reynolds number. We observe that numerical and experimental data are accurately described by a unique quadratic $D(h)$ spectrum which is found to extend from $h_{min} \approx 0.18$ to $h_{max} \approx 1$, as the signature of the highly intermittent nature of Lagrangian velocity fluctuations. 
\end{abstract}

\maketitle

Statistical properties of homogeneous three dimensional turbulence have been studied for a long time in the Eulerian framework \cite{FrischBook}. Recently a growing interest in studying intermittency from a dynamical point of view has been motivated by high precision Lagrangian experiments. Essentially two experimental groups have performed particle tracking in highly turbulent flows. The group at Cornell \cite{JFMBoden} reports measurements of Lagrangian acceleration in a turbulent water flow between two counter-rotating disks for Taylor-based Reynolds numbers $200 <  R_{\lambda} < 900$.  The experiment carried out at ENS-Lyon  \cite{MordantPint}, in a similar von K\'{a}rm\'{a}n flow,  is based on acoustic tracking. It provides Lagrangian velocity records covering the inertial range of turbulent motion, up to several integral time scales. In addition to these complementary experiments, DNS of the Navier-Stokes equations \cite{MordantPint,PopeYeung} have produced comparative numerical results in the range $75 < R_\lambda < 380$. Both experimental and numerical studies have revealed the existence of a very strong intermittency in the Lagrangian dynamics: the PDF of the velocity is Gaussian, while the PDF of the acceleration exhibits extremely large tails. In between, the PDFs of velocity  increments changes continuoulsy between these two functional forms as the time lag is decreased from integral times (the Lagrangian auto-correlation time) to dissipative ones (below the Kolmogorov time scale). Several stochastic models have been proposed that reproduce the behavior of the Lagrangian acceleration \cite{SawfordPOF} and of the Lagrangian velocity increments \cite{Rey03}. They rely on different physical assumptions and all include several {\it ad-hoc} hypothesis and parameters in order to fit the experimental observations --- a rather comprehensive review can be found in \cite{AriMaz03}. The aim of the present work is not to add to these models but to provide a comprehensive {\it description} of the Lagrangian intermittency, using a formalism that describes both the inertial and dissipative range of time scales. It is motivated by the desire to analyze globally the Lagrangian scales, which are likely to be dynamically connected (for example it is shown in \cite{MordantPOF} that the Lagrangian acceleration is influenced by the large scale dynamics, a feature that is also pointed out by models based on rapid distortion ideas \cite{LavDubNaz01}). In addition we require that the description of intermittency be independent on model assumptions. The multifractal description \cite{FrischParisi}, already widely used in Eulerian studies of turbulence, is a natural choice. To encompass inertial and dissipative features, we recast the multifractal picture of the intermediate dissipative range, originally proposed for Eulerian velocity fluctuations \cite{PalVul87,Men96},  into the context of Lagrangian velocity. The resulting analysis provides a synthetic and comprehensive description of the various sets of experimental and numerical  data.

In the present description, a first-order Lagrangian velocity increment over a time scale $\tau$ is written as :
\begin{equation}\label{eq1}
\delta_\tau v(t) = v(t+\tau)-v(t) = \beta\left(\tau/T\right)\delta_T v \ ,
\end{equation}
where all the time scale dependence is contained in the independent random function $\beta(\tau/T)$ ($>0$ since Lagrangian velocity increment PDFs are symmetric), $T$ being the integral (Lagrangian) time scale above which velocity increments become uncorrelated. The PDF of integral time scale increments $\delta_T v$ is thus assumed to be Gaussian ($\mathcal G$) --- a result of a central limit argument, also in agreement with Eulerian observations. Once the distribution of $\mathcal P(\beta)$ is known, the PDF of increments at any time scale $\tau$ is computed as
\begin{equation}\label{eq2}
\mathcal P (\delta_\tau v)  = \int \frac{d\beta}{\beta} \, \mathcal G \left(\frac{\delta_\tau v}{\beta}\right) \mathcal P(\beta)  \ .
\end{equation}
In the standard mutifractal formalism~\cite{FrischParisi}, $\beta$ is assumed to have a power law scale dependence in the inertial range, $\beta \sim (\tau/T)^h$, with a spectrum $\mathcal D(h)$ (meaning that the PDF of observing an exponent $h$ at scale $\tau$ is proportional to $(\tau/T)^{1-\mathcal D(h)}$). In the inviscid limit, $h$ and $\mathcal D(h)$ acquire the mathematical status of H$\ddot{\mbox{o}}$lder exponent and singularity spectrum, respectively. Note that this  description based on first order increments is restricted to exponents $h<1$ \cite{MuzBacArn93}, a limitation which will be addressed later. To describe  the entire range of scales covered in experimental measurements and computer simulations, one must take into account the effects of viscosity (finite $R_\lambda$). In the dissipative range, velocity fluctuations are smoothed by viscous damping (or by measurement filtering) and the velocity increments become proportional to the time scale $\delta _\tau v(t)  = \tau a(t)$, where $a(t)$ is the Lagrangian acceleration. We shall consider that the cross-over between inertial and dissipative statistics occurs when the local Reynolds number is of order unity, 
\begin{equation}
Re(\tau/T)  = \frac{\tau}{T} \, \beta^2 \left(\frac{\tau}{T} \right) \,  Re = 1 \ ,
\end{equation}
where $Re$ is the integral scale Reynolds number. This defines a local Kolmogorov dissipative time $\tau_\eta(h) = T Re^{\frac{-1}{2h+1}}$ where the local velocity increments change from inertial scale invariance to dissipative scaling (this implies $h \ge -1/2$, for time scales to be shorter than $T$). Changing the integration variable from $\beta$ to $h$ in Eq.(2), the PDF of velocity increments at any scale $\tau/T$ can be written as the sum of two contributions
\begin{eqnarray}
\mathcal P(\delta _\tau v)  = \int _{-1/2}^{h^*\left(\frac{\tau}{T}, Re\right)} dh \,  \frac{\mathcal P_i \left(h,\frac{\tau}{T},\mathcal D(h) \right)}{\beta_i \left(\frac{\tau}{T},h\right)} \,  \mathcal G\left(\frac{\delta _\tau v}{\beta_i \left(\frac{\tau}{T},h\right)}\right) \nonumber \\
+\int _{h^*\left(\frac{\tau}{T}, Re\right)}^{+\infty} dh \, \frac{\mathcal P_d  \left(h, Re,\mathcal D(h) \right)}{\beta_d \left(\frac{\tau}{T},h, Re\right)} \, \mathcal G\left(\frac{\delta _\tau v}{\beta_d \left(\frac{\tau}{T},h, Re\right)} \right) 
\end{eqnarray}
where the functions $\beta_{i,d}$ and $\mathcal P_{i,d}$ have the proper inertial ($\beta_i \sim (\tau/T)^h, \; \mathcal P_i \sim (\tau/T)^{1-\mathcal D(h)}$) and dissipative ($\beta_d \sim \tau/T, \; \mathcal P_d \sim (\tau_\eta(h)/T)^{1-\mathcal D(h)}$) scalings. The change occurs at the critical value $h^*$ for which the local Reynolds number is unity:
\begin{equation}
h^*\left(\frac{\tau}{T}, Re\right) = -\frac{1}{2}\left( 1+\frac{\ln  Re}{\ln \frac{\tau}{T}}\right) \mbox{ .}
\end{equation}
For $h < h^*(\tau/T, R_e)$, the increments $\delta _\tau v$ are in the inertial range, while they lie in the dissipative range for $h>h^*$.  Finally, we impose that the function $\beta(\tau/T)$ be continuous and differentiable at the transition, following a strategy used in the Eulerian domain~\cite{Men96}, and inspired from an elegant interpolation formula proposed by Batchelor \cite{Bat51}. In this framework, a single function $\beta(h; \tau/T, Re)$ covers the entire range of scale
\begin{equation}\label{eq11}
\beta \left(h; \frac{\tau}{T}, Re\right) = \frac{\left(\frac{\tau}{T}\right)^h}{\left( 1+ \left(\frac{\tau}{\tau_\eta(h)}\right)^{-\delta}\right)^{\frac{1-h}{\delta}}} \mbox{ ,}
\end{equation}
with probability density function
\begin{equation}\label{eq12}
\mathcal P \left(h;  \frac{\tau}{T}, Re, \mathcal D(h)\right) \sim \frac{\left(\frac{\tau}{T}\right)^{1-\mathcal D(h)}}{\left( 1+ \left(\frac{\tau}{\tau_\eta(h)}\right)^{-\delta}\right)^{\frac{\mathcal D(h)-1}{\delta}}} \ .
\end{equation}

The PDFs of Lagrangian velocity increments  are then computed using Eqs.(4-7), and compared to the experimental / numerical data. The entire range of time scales is covered when the four parameters $T, Re, \delta$ and $\mathcal D(h)$ are prescribed. $T$ and $Re$, both global parameters imposed by the flow production, are explicitly incorporated. $\mathcal D(h)$ is a parameter function which must be derived from the experimental or numerical data. In this work, and as {\it a posteriori} justified, we assume a quadratic form $\mathcal D(h) =1 - (h-c_1)^2 / 2c_2$ (a lower order approximation customary in intermittency studies). The constants $c_1$ and $c_2$ are constrained to satisfy $c_1= 1/2 + c_2$ in order for Kolmogorov scaling to be observed in the inertial range ($\langle \delta_\tau v ^2 \rangle \propto  \tau$). $\delta$ is a free parameter that accounts for the smoothing of the transition from inertial to dissipative scales. In his modeling of Eulerian velocity structure function, Batchelor \cite{Bat51} sets $\delta=2$, a value used in other Eulerian studies~\cite{Men96} and supported by theoretical arguments \cite{SirYakSmi94}. In practice $T$ and $Re$ are computed using the experimental data (up to some empirical multiplicative constants), and the free parameters $c_2$ and $\delta$ are estimated using a least square minimization scheme.

\begin{figure}[ht]
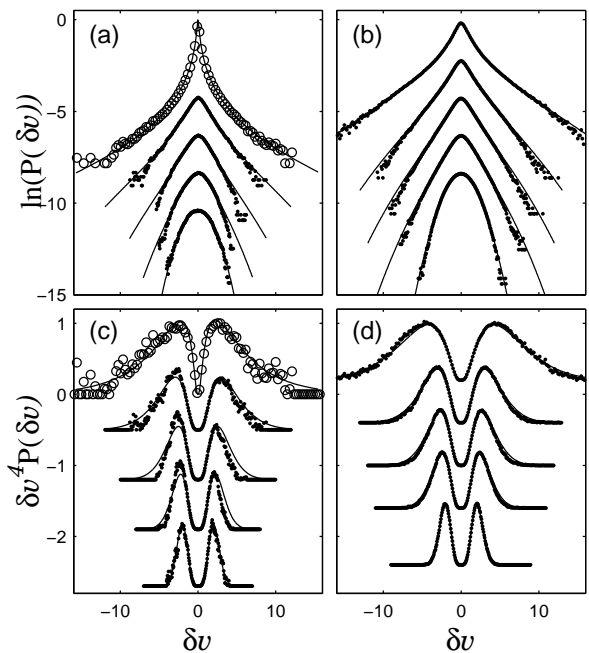

\figpdf
\caption{Comparison of the experimental (a,c) and numerical (b,d) data for the normalized velocity increment PDF $\mathcal P(\overline{\delta _\tau v})$, where $\overline{\delta _\tau v} = \delta _\tau v/\langle (\delta _\tau v)^2\rangle^{\frac{1}{2}}$, with the predictions of the multifractal description. 
($\bullet$)(a,c): ENS-Lyon experimental data, for time lags $\tau /T$ = 1, 0.35, 0.16 and 0.07, from bottom to top; the solid lines are the model fit with $c_2=0.075$ and $\delta=1.08$. 
($\bullet$)(b,d)  DNS data calculated for $\tau /T$ = 1, 0.25, 0.17, 0.11 and 0.05, from bottom to top; the solid lines correspond to parameter values  $c_2=0.086$ and $\delta=1.98$. 
($\circ$)(a,c) Cornell acceleration data, the solid lines are the model predictions for $c_2=0.079$ and $\delta=1.3$.  The curves are displayed with an arbitrary vertical shift for clarity, 
and the original $\delta v$-axis for the acceleration PDF ($\circ$)
has been shrunk by a factor of 4.}
\label{fig:pdf} 
\end{figure}

In Fig.\ref{fig:pdf}(a) we show the results for the experimental data from the ENS-Lyon group ($ R_{\lambda} = 740$) \cite{MordantPint}. The velocity increment PDFs are well reproduced at all scales for the set of parameter values $\delta=1.08$, $c_1=0.575$ and $c_2=0.075$. This agreement is emphasized in Fig.\ref{fig:pdf}(c), where the fourth order moment $(\overline{\delta _\tau v})^4\mathcal P(\overline{\delta _\tau v})$ is shown to peak at the same values as the experimental curves and this for all considered time scales. In Figs.\ref{fig:pdf}(a,c) we have also included  the Cornell  acceleration data at $R_{\lambda} = 690$ \cite{JFMBoden}. What is remarkable is that with quite consistent parameter values, namely $\delta=1.3$ and still a parabolic $\mathcal D(h)$ curve with $c_1=0.579$ and $c_2=0.079$, one again reproduces the experimental acceleration PDF with a great accuracy in the tails; in particular, the flatness  $F=\langle a^4\rangle/\langle a^2\rangle^2 = 56.1$ is very close to the experimental value $F=55 \pm 8$ \cite{JFMBoden}. For our DNS data at $R_{\lambda} = 140$ (Figs.\ref{fig:pdf}(b,d)), the agreement is even more pronounced than for the experimental data. In particular, it is hard to distinguish the fit from the data points in Fig.\ref{fig:pdf}(d). Note that the parameter values obtained to get the best fit of the numerical data are not significantly different from those used in Figs.\ref{fig:pdf}(a,c): $\mathcal D(h)$ corresponds to $c_1=0.586$ and $c_2=0.086$, while $\delta = 1.98$ is now quite close to the Batchelor value $\delta=2$ \cite{Bat51}. The fact that $\delta$ is smaller in experimental measurements is due to the filtering induced by the finite size of the tracer particle and signal processing algorithms.

\begin{figure}
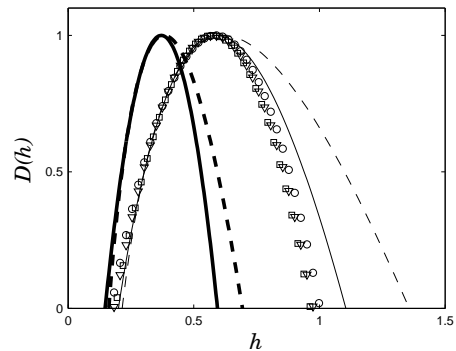

\figdh
\caption{\label{fig:dh} $\mathcal D(h)$ curves extracted from: ($\square$) ENS-Lyon velocity data, ($\nabla$) Cornell acceleration data and ($\circ$) DNS numerical data. Are also represented for comparison the Eulerian log-normal (thick solide line) and log-Poisson (thick dashed line) spectra as well as their Lagrangian counter-parts   computed using Eq.(9) (solid and dashed lines respectively).  }
\end{figure}

A first finding of our analysis is that almost identical functions $\mathcal D(h)$ are obtained for the three sets of data (Cornell, Lyon and DNS), although they cover a wide range of scales and of turbulent Reynolds numbers: the symbols in Fig.\ref{fig:dh} are undistiguishable, certainly within error bars. This suggests that such a parabolic $\mathcal D(h)$ curve should also be relevant in the limit of infinite Reynolds number. A parabolic singularity spectrum is the hallmark of log-normal statistics as originally proposed by Kolmogorov and Obukhov \cite{Kol62} for Eulerian velocity statistics. Our findings for Lagrangian velocity statistics is a parabolic $\mathcal D(h)$ spectrum centered at $c_1=0.58\pm 0.01$, significantly larger than the K41 value $1/2$, and of width $c_2 = 0.08 \pm 0.01$ (commonly called the intermittency exponent). This is significantly larger than the corresponding value $c_2^E = 0.025 \pm 0.003$ derived for Eulerian velocity data \cite{DelMuzArn01}. It corroborates the fact that Lagrangian velocity statistics are more intermittent than Eulerian velocity statistics~\cite{MordantPint}. Let us remark that the ratio of the Lagrangian to Eulerian intermittency exponents $c_2 / c_2^E = 0.08/0.025 = 3.2 \pm 0.2$ is very close to the value $\left(3/2\right)^3 = 3.375$ which can be predicted using a Kolmogorov-Richardson argument  \cite{MordantPint,Bor93}. 

\begin{figure}[t!]
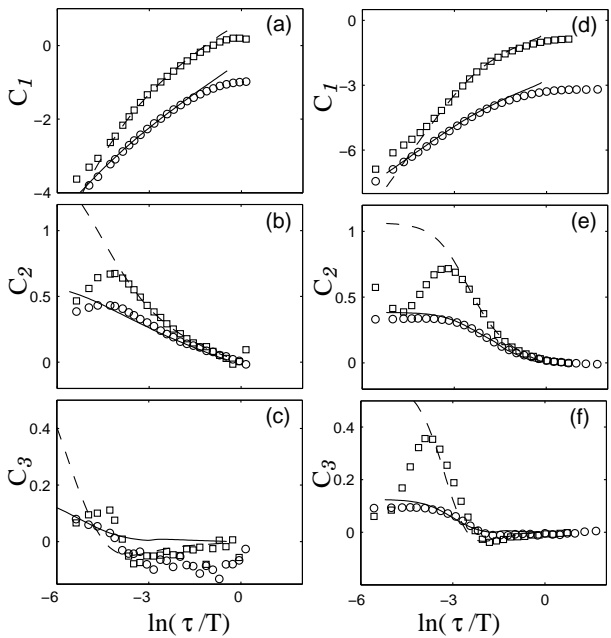

\figcum
\caption{
Cumulant analysis $C_n\left(\tau/T\right)$ vs $\ln \tau/T$ for: (a-c) ENS-Lyon  experimental data and (d-f) DNS data. ($\circ$) first-order velocity increments ($N=1$); ($\square$) second-order velocity increments ($N=2$). The associated multifractal descriptions for $N=1$ (solid line) and $N=2$ (dashed line) correspond to the same parameter values as in Fig.1. Curves in (a,d) are vertically shifted for clarity.} 
\label{fig:cumul}
\end{figure}

To emphasize further the quality of the proposed description of Lagrangian intermittency, we now consider the evolution across the scales of the moments of the velocity increments, the so-called structure functions \cite{FrischBook}, $S(p,\tau) = <|\delta_\tau v|^p>$. As advocated in Ref.\cite{DelMuzArn01} for Eulerian velocity data analysis, the magnitude cumulant analysis provides a more reliable alternative to the structure function method. Indeed, it is straightforward to derive the relationship between the moments of $\mathcal P(|\delta _\tau v|)$ and the cumulants $C_n(\tau)$ of $\mathcal P(\ln |\delta _\tau v|)$:
\begin{equation}\label{eq15}
<|\delta _\tau v|^p> =\exp \left( \sum_n C_n(\tau)\frac{p^n}{n!}\right) \ . 
\end{equation}
In the inertial range, where multifractal power law scaling is expected to be observed for the structure functions, the cumulants should behave like $C_n(\tau) \sim c_n\ln \tau$. In Fig.\ref{fig:cumul}, we report the results of the computation of the cumulant $C_n(\tau)$ for both the ENS-Lyon experimental data (Figs \ref{fig:cumul}(a-c)) and our DNS numerical data (Figs \ref{fig:cumul}(d-f)). We have explicitly subtracted the cumulant of the $\delta_T v$ Gaussian PDF and the curves are computed using the same parameter set as in Fig.\ref{fig:pdf}. For both the experimental and numerical cumulants, it is quite convincing that our multifractal decription provides a comprehensive understanding of the observed departure from scaling when going from large $\tau/T \sim 1$ to small $\tau/T\sim 10^{-2}$ time scales. In fact, an unambiguous inertial range is observed for the experimental second order cumulant only (Fig.\ref{fig:cumul}(b)). For the first order cumulant, both the experimental (Fig.\ref{fig:cumul}(a)) and numerical (Fig.\ref{fig:cumul}(d)) curves display some curvature over the entire range of scale, a feature that is well reproduced by our extended multifractal description.  For $C_2(\tau/T)$, the model describes quite well the cross-over behavior observed in the experimental (Fig.\ref{fig:cumul}(b)) and numerical (Fig.\ref{fig:cumul}(e)) data down to time scales $\tau/T$ of the order of the smoothing filtering scale (including the finite size of the tracer particles). The predicted plateau $C_2(\tau/T)\rightarrow \overline{C}_2 (R_{\lambda})$ in the limit $\tau/T\rightarrow 0$, is reached in the numerical but not in the experimental data. For $C_3(\tau/T)$, we note that the DNS data in the inertial range are quite well reproduced taking $c_3=0$,  at odds with the experiment. 

We now return to the assumption that the scaling exponent $h$ should remain smaller than one. It may be seen in Fig.2 that the largest measured $h$ values reach 1. In this case, it is known that first order increments are not adapted to detect singularities with exponents $h \ge 1$~\cite{MuzBacArn93}. We have thus repeated our analysis using 
second order velocity increments $\delta_\tau ^{(2)} v(t) = v(t+\tau)-2v(t+\tau/2)+v(t)$. Eqs.(4,6,7) must be slightly modified because (i) the dissipative range should now scale as $(\tau/T)^2$ and (ii) $h^*$ may be larger than one and the integrals must be computed in the order of increasing $h$'s. When using the same parameters as before, we observe in Fig.\ref{fig:cumul} that the experimental and numerical cumulants are robustly and even better reproduced. Let us point out that as seen in Figs. \ref{fig:cumul}(b,c,e,f), because the $C_n(\tau/T)$ for $n\ge 2$ are predicted to reach a plateau $\overline{C}_n(R_{\lambda},N) \sim (1+2N)^n \ln R_\lambda$ in the limit of vanishing $\tau$'s, then the experimental and numerical cumulants for the second-order increments are much more affected by the smoothing filtering process in the intermediate dissipative range than previously observed for the first-order increments. 

To conclude, we return to our observation that a unique $\mathcal D(h)$ spectrum yields an accurate description of the Lagrangian velocity statistics at all scales. Such a spectrum $\mathcal D^E(h)$ has been extensively studied in the Eulerian domain \cite{FrischBook}. Two widely used forms (corresponding to log-normal and log-Poisson statistics) are shown in Fig.2. They can be mapped into the Lagrangian domain
\begin{equation}
\mathcal D(h) = -h + (1+h) \mathcal D^E \left( h/(1+h) \right) \ ,
\end{equation} 
using a Kolmogorv Refined Similarity argument in the spirit of the work done by Borgas \cite{Bor93} . The resulting curves are shown in Fig.2; we note that the agreement with the measured Lagrangian $\mathcal D(h)$ functions is excellent on the left-hand side of the curves, {\it i.e.} for values $h < c_1$ corresponding to intense velocity increments. On the right-hand side ($h > c_1$) there is a noticeable difference. Whether this difference is significant deserves more investigation. It may be of importance since the above relationship clearly shows that the Eulerien and Lagrangian singularity spectra cannot be both log-normal.

Numerical simulations were performed at CINES (France) using an IBM SP computer. We wish to acknowledge Bernard Castaing for his critical comments. 
We are grateful to the Cornell team for making their data available. 


\end{document}